\documentclass[reprint,amsmath,amssymb,aps]{revtex4-1}

\usepackage{graphicx}
\usepackage{subfigure}
\usepackage{dcolumn}
\usepackage{bm}
\usepackage{bbold}

\usepackage{color,soul}

\begin{document}
\title{Rotation-invariant observables as Density Matrix invariants}

\author{Margarita Gavrilova}
\email{gavrilova@phystech.edu}
\affiliation{JINR, 141980 Dubna, Russia}
\affiliation{MIPT, 141700 Dolgoprudny, Russia}
\author{Oleg Teryaev}%
\email{teryaev@jinr.ru}
\affiliation{JINR, 141980 Dubna, Russia}

\date{\today}

\begin{abstract}
It is shown that general dilepton angular distribution (with parity violating terms taking into account) in vector particle decays can be described through a set of five $SO(3)$ rotational-invariant observables. These observables are derived as invariants of the spacial part of the hadronic tensor (density matrix) expressed in terms of angular coefficients. The restrictions on the invariants following from the positivity of the hadronic tensor are obtained. Special cases of $SO(2)$ rotations are considered. Calculation of invariants for available data on $Z$ and $J/\psi$ decays is performed.
\end{abstract}

\maketitle

\section{\label{sec:level1}Introduction}
The Drell-Yan-type processes in which a lepton pair is produced in hadronic collisions are the sensitive tests  of Standard Model and probes of New Physics. The precision measurements of dilepton angular coefficients at various energies were recently presented by CDF~\cite{CDF}, CMS~\cite{CMS} and ATLAS~\cite{ATLAS} collaborations (for $Z$ decays) and by PHENIX~\cite{adare2017angular} collaboration (for $J/\psi$ decays). As the values of angular coefficients depend on the choice of a reference frame, an adequate comparison between observables measured in different coordinate systems (and between theory and experiment)  
 may be performed for frame-independent quantities. Such quantities provide a powerful tool for the data analysis and can reveal systematic biases that were not taken into account.

The great progress was achieved in this direction. Several invariants for special $SO(2)$ rotations were proposed~\cite{faccioli2010rotation, faccioli2010rotationPV, palestini2011angular, faccioli2011model} and even a general recipe for constructing $SO(3)$ conserving parameters was recently developed~\cite{ma2017rotation}.

In this work, we suggest a method which allows to find frame-independent quantities for vector particle (like virtual photons or electroweak bosons) decays. We had two main motivations. First, to construct such a procedure which would give tools not only to reproduce previous results but also to constrain them. Second, to simplify known invariants whenever possible. The key idea of proposed method is to express the hadronic tensor corresponding to the process, which also happens to be initial state density matrix, in terms of coefficients of final state dilepton angular distribution. This procedure was proposed and realized for a case of parity-conserving angular distribution in~\cite{teryaev2011positivity}. As we discuss later in the text, in the center of mass frame the tensor reduces to $3 \times 3$ hermitian matrix and we can focus on studying the invariants of this matrix. It is well known, that invariants of a matrix always can be written through eigenvalues. Once this is done, we apply positivity conditions to bound the invariants.

In what follows we express the hadronic tensor in terms of observables, consider its irreducible representations and show that symmetric and antisymmetric parts as well as various combinations of them provide us with invariants of the angular distributions. We obtain five $SO(3)$ rotational invariants and relate them to previously proposed invariant parameters~\cite{ma2017rotation}. Then we explicitly write restrictions on invariants using the positivity of the hadronic tensor and normalization condition. In the later section, we show that additional invariants appear when $SO(2)$ rotation around a fixed axis is considered. Finally, we present a calculation of invariants for data on $Z$ decays released by ATLAS~\cite{ATLAS} and data on $J/\psi$ decays by PHENIX~\cite{adare2017angular} collaboration. For the later we also consider a geometric model \cite{OT05, teryaev2011positivity} interpretation, which later allowed, by including the additional concept of non-coplanarity angle,  to describe also the violation of  Lam-Tung relations and classify the rotational invariants\cite{peng2016interpretation, Peng:2018tty,  Chang:2018pvk} for Drell-Yan and quarkonium production in both collider and fixed-target experiments.

\section{\label{sec:level1}General form of angular distribution}
In our study, we consider an annihilation process via the vector particle which in its turn decays into a lepton pair. For such a process the hadronic tensor can be written, following \cite{ teryaev2011positivity} in terms of spin observables (the coefficients of the angular distribution). To do this first we should consider the general form of the angular distribution. Two parametrization~\eqref{general_lambda} and~\eqref{general_A} are widely used in literature. The first one usually appears in theoretical papers (see for instance~\cite{faccioli2010rotation, faccioli2010rotationPV, palestini2011angular, faccioli2011model, ma2017rotation}), while the second one can be found in experimental works (see~\cite{CDF, CMS, ATLAS}):
\begin{eqnarray}\label{general_lambda}
    \frac{1}{\sigma}\frac{\text{d} \sigma}{\text{d}\Omega} & = & \frac{3}{4 \pi} \frac{1}{3 + \lambda_\theta} \left( 1     + \lambda_\theta \cos^2 \theta + \lambda_{\theta \phi} \sin 2 \theta \cos \phi \right. \nonumber \\
    & + & \lambda_\phi \sin^2 \theta         \cos 2\phi +         \lambda_{\perp \phi}\sin^2 \theta \sin 2\phi \nonumber \\
    & + &\lambda_{\perp\theta\phi} \sin 2\theta \sin \phi + 2 A_\theta \cos \theta \nonumber \\
    & + & \left. 2 A_\phi \sin \theta \cos \phi + 2 A_{\perp \phi} \sin \theta \sin \phi\right),
\end{eqnarray}
\begin{eqnarray}\label{general_A}
    \frac{1}{\sigma}\frac{\text{d}\sigma}{\text{d}\Omega} & = & \frac{3}{16 \pi}\left(\left(1 + \frac{A_0}{2}\right) + \left(1-\frac{3}{2}A_0\right)\cos^2\theta \right. \nonumber \\
    & + & A_1\sin2\theta\cos\phi + \frac{A_2}{2}\sin^2\theta\cos2\phi \nonumber \\
    & + & A_3\sin\theta\cos\phi +A_4\cos\theta + A_5\sin^2\theta\sin2\phi \nonumber \\
    & + & \left. A_6\sin 2\theta\sin\phi + A_7\sin\theta\sin\phi\right).
\end{eqnarray}

Before moving to the derivation of the hadronic tensor let us first discuss the relation between two parametrizations. Both of them are chosen in the way which ensures that the total cross section
\begin{equation}\label{s_total}
    \sigma_\text{total} = \int \frac{\text{d} \sigma}{\text{d}\Omega} \text{d}\Omega
\end{equation}
is a constant (there is no parameter dependence). In~\eqref{general_lambda} this is guaranteed by the common factor $1/(3 + \lambda_\theta)$. Yet the parametrization~\eqref{general_A} is more convenient because with it angular coefficients do not occur in the total cross section even without common factor with one of the coefficients in the denominator. Because of this the hadronic tensor takes a simpler form when written in terms of these parameters.

Numerical factors $3/4\pi$ in~\eqref{general_lambda} and $3/16 \pi$ in~\eqref{general_A} are often omitted in literature, however, if one wants to find the relation between two sets of parameters, factors are important since they ensure the equality of two angular distributions~\eqref{general_lambda} and~\eqref{general_A} and corresponding total crosssections~\eqref{s_total}. Comparing the coefficients one can express parameters in~\eqref{general_lambda} through parameters in~\eqref{general_A} and vice versa:
\begin{eqnarray}\label{A-l_relation}
    \lambda_\theta = \frac{2-3 A_0}{2 + A_0},\quad A_0 = \frac{2 \left(1 - \lambda _{\theta }\right)}{3 + \lambda_\theta} \nonumber \\
    \lambda_\phi = \frac{A_2}{2 + A_0},\quad A_2 = \frac{8 \lambda _{\phi }}{3 + \lambda_\theta}\nonumber \\
    \lambda_{\theta \phi} = \frac{2 A_1}{2 + A_0},\quad A_1 = \frac{4 \lambda _{\theta \phi }}{3 + \lambda_\theta}\nonumber \\
    \lambda_{\perp \phi} = \frac{2 A_5}{2 + A_0}, \quad A_5 = \frac{4 \lambda_{\perp \phi}}{3 + \lambda_\theta}\nonumber \\
    \lambda_{\perp \theta \phi} = \frac{2 A_6}{2 + A_0}, \quad A_6 = \frac{4 \lambda_{\perp \theta \phi}}{3 + \lambda_\theta}\nonumber \\
    A_\theta = \frac{A_4}{2 + A_0}, \quad A_4 = \frac{8 A_\theta}{3 + \lambda_\theta}\nonumber \\
    A_\phi = \frac{A_3}{2 + A_0}, \quad A_3 = \frac{8 A_\phi}{3 + \lambda_\theta}\nonumber \\
    A_{\perp \phi} = \frac{A_7}{2 + A_0}, \quad A_7 = \frac{8 A_{\perp \phi}}{3 + \lambda_\theta}
\end{eqnarray}  
Once the relation between parameters is found the numerical factors in~\eqref{general_lambda} and~\eqref{general_A} are not important provided that only one of parameterizations is used. We consider parametrization~\eqref{general_lambda} in our analysis and omit the corresponding factor everywhere. One can easily switch to another parametrization at any stage using relations~\eqref{A-l_relation}.

\section{\label{sec:level1}Hadronic tensor in terms of observables}
Now after the above opening remarks we can move to derivation of the hadronic tensor through observables. To do this at first we should recall that cross section is proportional to the contraction of hadronic $W^{\mu \nu}$ and leptonic $L^{\mu \nu}$ tensors:
    \begin{equation}
        \frac{1}{\sigma}\frac{\text{d} \sigma}{\text{d}\Omega} \propto W^{\mu \nu} L_{\mu \nu}.
    \end{equation}
Let us work in the dilepton rest frame. Due to conservation of quark currents, the hadronic tensor satisfy the transversity condition $q_\nu W^{\mu\nu} = 0$, where $q_\nu$ is 4-momenta of the intermediate vector particle~\cite{drell1992theoretical}. In the reference frame under consideration
\begin{equation}
    q_\nu = (q_0, 0, 0, 0)
\end{equation}
and thus only a spatial part of the tensor can be non-zero. Therefore, the hadronic tensor reduces to hermitian $3 \times 3$ matrix:
    \begin{equation}\label{h_tensor}
    W^{i j} = \left(\begin{array}{ccc}
        d_1 & a_1 + i a_2 & b_1 + i b_2 \\
        a_1 - i a_2 & d_2 & c_1 + ic_2 \\
        b_1 - i b_2 & c_1 - i c_2 & d_3\\
\end{array}\right),
    \end{equation}
where $a_{1,2}$, $b_{1,2}$, $c_{1,2}$ and $d_{1,2,3}$ are real parameters.

If negligibly small lepton masses are assumed, the spatial part of leptonic tensor takes the following form~\cite{drell1992theoretical}:
    \begin{equation}\label{l_tensor}
        L_{i j} \propto \delta_{i j} - n_i n_j + i g \epsilon_{ijk}n^k
    \end{equation}
where $n_i$ is a vector parallel to the momentum of the final state positively charged lepton
    \begin{equation}\label{n}
        \bar{n} = \left( \cos \theta, \sin \theta \cos \phi, \sin \theta \sin \phi\right)
    \end{equation}
and $g$ is a numerical constant.

Note that symmetric $\delta_{i j} - n_i n_j$ and antisymmetric part $i \epsilon_{ijk}n^k$ enter $L_{ij}$ with different prefactors in general case. For example, explicit calculation for $Z$-decay into a lepton pair gives factors $(c_a^2 + c_v^2)$ and $2 c_a c_v$ respectively
\begin{equation}
    g = \frac{2 c_a c_v}{c_a^2 + c_v^2},
\end{equation}
where $c_a$ and $c_v$ are axial and vector constants. However, it occurs that $g$ doesn't affect the invariance of relations we are aimed to find. It only enters invariants as a common factor, which is not important. One would be able to see this later from the form of invariants. Thus we can safely choose $g = 1$.
    
Contracting the leptonic tensor~\eqref{l_tensor} with the hadronic tensor in the form of~\eqref{h_tensor} and setting $g = 1$ we obtain the angular distribution, but with angular coefficients written in terms of $a$, $b$, $c$ and $d$.
\begin{eqnarray}
    \frac{1}{\sigma}\frac{\text{d}\sigma}{\text{d}\Omega} & \propto & (d_1 + \frac{1}{2}d_2 + \frac{1}{2}        d_3) + (-d_1 + \frac{1}{2}d_2 + \frac{1}{2}d_3)            \cos^2\theta \nonumber \\
    & - & a_1\sin2\theta\cos\phi + \frac{1}{2}(d_3 - d_2)\sin^2\theta\cos2\phi \nonumber \\  
    & - & b_1\sin 2\theta\sin\phi - c_1\sin^2\theta\sin2\phi \nonumber \\
    & - & 2a_2\sin\theta\sin\phi     + 2b_2\sin\theta\cos\phi - 2c_2\cos\theta
\end{eqnarray}
Expressing those parameters through the coefficients introduced in~\eqref{general_lambda} we get the hadronic tensor $W^{ij}$
\begin{equation}\label{h_tensor_general_lambda}
\frac{2}{3+\lambda_\theta}\begin{pmatrix}
        \frac{1 - \lambda_\theta}{2} & -\lambda_{\theta \phi} - i A_{\perp \phi}  & -\lambda_{\perp \theta \phi} + i A_\phi\\
        -\lambda_{\theta \phi} +  i A_{\perp \phi}&\frac{ 1 + \lambda_\theta - 2 \lambda_\phi}{2} & -\lambda_{\perp \phi} - i A_\theta\\
        -\lambda_{\perp \theta \phi} - i A_\phi & -\lambda_{\perp \phi} + i A_\theta & \frac{1 + \lambda_\theta + 2\lambda_\phi}{2}\\
\end{pmatrix}
\end{equation}
Here the normalization condition $\text{Tr} \hspace{3pt} W = 1$ is imposed. This is the first important result -- we are left with hadronic tensor expressed in terms of spin observables. Hadronic tensor written in the form of~\eqref{h_tensor_general_lambda} is a generalization of the previous result appeared in~\cite{teryaev2011positivity} to the case of the most general angular distribution when antisymmetric terms are also taken into account. This matrix contains all the information about the angular distribution. Therefore, in this formalism the problem of searching for the frame independent invariants is equivalent to the search for the invariants of the matrix~\eqref{h_tensor_general_lambda}.

\section{\label{sec:level1}Invariants}

According to~\cite{ma2017rotation} one should expect the presence of $8 - 3 = 5$ independent rotational invariants, where $8$ is a number of parameters in the distribution and $3$ correspond to three Euler angles used to parameterize an arbitrary $SO(3)$ rotation. This counting can be understood by the following geometric picture. Let us consider the eight dimensional parameter space. Particular angular distribution in the fixed coordinate frame can be identified with a point in this space. Different coordinate frames are related by rotation described by three parameters, thus the set of points corresponding to the particular distribution, but written in all the possible coordinates will correspond to three dimensional hypersurface. To describe a $d$-dimensional hypersurface in $D$-dimensional space one needs $D - d$ independent equations, for example, in a form $f_i (A_0, ...) = 0, \text{where } i = 1, .., D - d$. Functions $f_i$ then provide a full set of independent invariants.

 In addition, a simple example of photon density matrix illustrates this reasoning. Photon density matrix can be expanded in terms of Pauli matrices with coefficients being three Stokes parameters $S_3, S_1, S_2$. Since photon is massless, it's polarization vector can be subjected to rotations only in the plane perpendicular to photon's momentum, that is why one expects to have two rotational invariants, which are well known quantities corresponding to the separation of the symmetric and asymmetric parts: $S_1^2 + S_3^2$ and $S_2^2$. It is interesting to note that this is not the case for spin--$\frac{1}{2}$ particles, where one considers three dimensional rotations instead of two dimensional ones and cannot explicitly separate symmetric and antisymmetric parts.

To find all five invariants of the angular distribution in the hadronic tensor formalism we decompose the hermitian matrix~\eqref{h_tensor_general_lambda} into a sum of unit trace-$1$ matrix, traceless symmetric matrix $W_s$ and traceless antisymmetric matrix $W_a$:
\begin{equation}\label{decomposition}
W = \frac{1}{3} \cdot \mathbb{1} + W_s + i W_a,
\end{equation}
where $\mathbb{1}$ denotes $3 \times 3$ unity matrix. The spacial part of the hadronic tensor transforms under an arbitrary frame rotation as follows
\begin{equation}\label{rotation_S}
    W^\prime = S^T W S,
\end{equation}
where $W^\prime$ is a hadronic tensor written in a new coordinate frame, $S$ is real orthogonal $3 \times 3$ matrix belonging to $SO(3)$ group. After applying transformation~\eqref{rotation_S} to decomposition~\eqref{decomposition} we obtain
\begin{equation}\label{rotation_S_trace}
    W^\prime = \frac{1}{3} \cdot \mathbb{1} + S^\text{T}\left( W_s + i W_a \right)S = \frac{1}{3} \cdot \mathbb{1} + S^T W_s S + i S^T W_a S.
\end{equation}
Matrices $W_s$ and $W_a$ transform independently, because of the reality of S. Expression~\eqref{rotation_S_trace} shows that invariants of matrices $W_s$, $W_a$, $W_s + i W_a$ and various combinations of them are also invariants of the total matrix $W$. Since eigenvalues are preserved by $SO(3)$ rotations and all invariants of a matrix can be written in terms of them, we are interested in finding five independent eigenvalues or combinations of eigenvalues for these matrices. They can be found as roots of characteristic equation which for arbitrary matrix $F$ with eigenvalues $f$ takes the form:
\begin{equation}
    \text{det}\left[ F - f \cdot \mathbb{1}\right] = 0,
\end{equation}
Below in~\eqref{eq} we list characteristic equations obtained for matrices $W_a$, $W_s$, $W_s + i W_a$ and $W_a W_s$ (which is the same as for $W_s W_a$) with eigenvalues denoted as $w^{(a)}$,  $w^{(s)}$, $w$ and $w^{(as)}$ respectively. Expressions for coefficients of characteristic equations in terms of angular parameters are given in~\eqref{notation}.
\begin{subequations}\label{eq}
\begin{equation}\label{eq_a}
    w^{(a)} \left( {w^{(a)}}^2 + 4 U_1 \right) = 0
\end{equation}
\begin{equation}\label{eq_s}
   {w^{(s)}}^3 -\frac{4}{3} U_2 {w^{(s)}} - \frac{8}{27} T = 0
\end{equation}
\begin{equation}\label{eq_tot}
    w^3 -\left(4U_1 + \frac{4}{3} U_2 \right) w -\frac{8}{27} (T + R) = 0
\end{equation}
\begin{equation}\label{eq_as}
\qquad w^{(as)} \left( {w^{(as)}}^2 + \frac{16}{9} M \right) = 0
\end{equation}
\end{subequations}
\begin{widetext}
\begin{subequations}\label{notation}
\begin{equation}\label{U1}
    U_1 = \frac{A_\theta^2 + A_\phi^2 + A_{\perp \theta \phi}^2}{(3 + \lambda_\theta)^2}, \qquad U_2 = \frac{\lambda_\theta^2 + 3\left( \lambda_\phi^2 + \lambda_{\theta \phi}^2 + \lambda_{\perp \phi}^2 + \lambda_{\perp \theta \phi}^2 \right)}{(3 + \lambda_\theta)^2}
\end{equation}
\begin{equation}\label{T}
T = \frac{\left(\lambda_\theta + 3 \lambda_\phi\right)\left(2 \lambda_\theta^2 -6 \lambda_\theta \lambda_\phi + 9 \lambda_{\theta \phi}^2\right) + 9 \left(\lambda_\theta \lambda_{\perp \theta \phi}^2 - 2 \lambda_\theta \lambda_{\perp \phi}^2 + 6 \lambda_{\theta \phi}\lambda_{\perp \theta \phi} \lambda_{\perp \phi} - 3 \lambda_\phi \lambda_{\perp \theta \phi}^2\right)}{\left(3 + \lambda_\theta\right)^3}
\end{equation}
\begin{equation}\label{R}
R = \frac{1}{\left(\lambda _{\theta }+3\right)^3}\left( 54 \left(A_{\theta } A_{\phi } \lambda _{\theta \phi }+A_{\theta } A_{\perp\phi} \lambda _{\perp\theta \phi}+A_{\perp\phi} A_{\phi } \lambda _{\perp\phi}\right)+9 \lambda _{\theta } \left(2 A_{\theta }^2-A_{\perp\phi}^2-A_{\phi }^2\right)+27 \lambda _{\phi } \left(A_{\phi }^2-A_{\perp\phi}^2\right) \right)
\end{equation}
\begin{eqnarray}\label{M}
    M & = & \frac{1}{(3 + \lambda_\theta)^4} \left \{ A_{\theta }^2 \left(\lambda _{\theta }^2-9 \lambda _{\phi }^2-9 \lambda _{\perp\phi}^2\right) - A_{\phi }^2 \left(2 \lambda _{\theta } \left(\lambda _{\theta }+3 \lambda _{\phi }\right)+9 \lambda _{\perp\theta \phi}^2\right) + A_{\perp\phi}^2 \left(6 \lambda _{\theta } \lambda _{\phi }-2 \lambda _{\theta }^2-9 \lambda _{\theta \phi }^2\right) \right. \nonumber \\
    & + & \left. 6 A_{\theta } A_{\perp\phi} \left(\lambda _{\perp\theta \phi } \left(\lambda _{\theta }-3 \lambda _{\phi }\right)+3 \lambda _{\theta \phi } \lambda _{\perp\phi}\right) + 6 A_{\phi } \left[A_{\theta } \left(\lambda _{\theta \phi } \left(\lambda _{\theta }+3 \lambda _{\phi }\right)+3 \lambda _{\perp\theta \phi} \lambda _{\perp\phi}\right)+A_{\perp\phi} \left(3 \lambda _{\theta \phi } \lambda _{\perp\theta \phi}-2 \lambda _{\theta } \lambda _{\perp\phi}\right)\right]
 \right \} \nonumber \\
\end{eqnarray}
\end{subequations}
\end{widetext}
Parameters $U_1$, $U_2$, $T$, $R$ and $M$ are all rotational invariants. Vieta's theorem~\cite{polyanin2006handbook} relates them to eigenvalues of matrices $W_a$, $W_s$, $W_s + i W_a$ and $W_a W_s$ by the expressions listed in~\eqref{nice_relations}, where indexes $1, 2, 3$ enumerate eigenvalues.
\begin{subequations}\label{nice_relations}
\begin{equation}\label{wa2}
    w_1^{(a)} w_2^{(a)} + w_1^{(a)} w_3^{(a)} + w_2^{(a)} w_3^{(a)} = w_2^{(a)} w_3^{(a)} = 4 U_1
\end{equation}
\begin{equation}\label{ws2}
w_1^{(s)} w_2^{(s)} + w_1^{(s)} w_3^{(s)} + w_2^{(s)} w_3^{(s)} =  -\frac{4}{3} U_2
\end{equation}
\begin{equation}\label{ws3}
w_1^{(s)} w_2^{(s)} w_3^{(s)}  = \frac{8}{27} T
\end{equation}
\begin{equation}\label{w2}
    w_1 w_2 + w_1 w_3 + w_2 w_3 = -4 U_1 - \frac{4}{3} U_2
\end{equation}
\begin{equation}\label{w3}
    w_1 w_2 w_3 = \frac{8}{27} \left(T + R\right)
\end{equation}
\begin{eqnarray}\label{was}
w_1^{(as)} w_2^{(as)} + w_1^{(as)} w_3^{(as)} + w_2^{(as)} w_3^{(as)} & = & w_2^{(as)} w_3^{(as)} \nonumber \\ & = & 4 M
\end{eqnarray}
\end{subequations}

The explicit expressions for eigenvalues are somewhat cumbersome (see Appendix) and for practical purposes it is more convenient to use combinations of them listed in~\eqref{notation},~\eqref{nice_relations}.

In~\eqref{expr} we express $SO(3)$ invariants derived in the work~\cite{ma2017rotation} in terms of parameters that we introduced in~\eqref{notation}. To avoid possible confusion, we use tilda sign to denote parameters from~\cite{ma2017rotation}. Note, that $U_1$ and $U_2$ are equal to the same-name invariants derived in~\cite{ma2017rotation} up to unimportant numerical factors. It is also worth mentioning that parameter $U_1$ comes from the antisymmetric part of the hadronic tensor, which can be written as $W_a = 2 \epsilon_{ijk} A_k$, where $\bar{A} = \frac{1}{3 + \lambda_\theta}(A_\theta, A_\phi, A_{\perp \phi})$. From this point of view $U_1$ is equal to $\bar{A}^2$, the squared length of the vector corresponding to the vector part of the density matrix.
\begin{subequations}\label{expr}
\begin{equation}
    \overset{\sim}{U}_1 = \frac{3}{\pi} U_1
\end{equation}
\begin{equation}
    \overset{\sim}{U}_2 = \frac{1}{5 \pi} U_2
\end{equation}
\begin{equation}
    \overset{\sim}{W}_3 = \frac{1}{70 \pi^2} \left(T + 7 R \right)
\end{equation}
\begin{eqnarray}
    \overset{\sim}{W}_4 & = & \frac{9}{20 \pi}\overset{\sim}{U}_1^2 + \frac{15}{28 \pi}\overset{\sim}{U}_2^2 + \frac{27}{14\pi}\overset{\sim}{U}_1 \overset{\sim}{U}_2 \nonumber \\
    & - & \frac{9}{35 \pi^3}\frac{1}{144}\left( 45 U_1 + 10 R + 36 M\right)
\end{eqnarray}
\begin{eqnarray}
    \overset{\sim}{W}_5  & = &  \frac{5}{2 \pi} \left( \frac{3}{7} \overset{\sim}{U}_1 + \frac{5}{11} \overset{\sim}{U}_2\right)\overset{\sim}{W}_3 \nonumber \\
    & + &  \frac{3}{539 \pi^4} \left( U_1 \left( -\frac{143}{4} + 429 U_2 - 297 T \right) + \frac{7}{3}U_2 R\right) \nonumber \\
\end{eqnarray}
\end{subequations}
Note that with help of~\eqref{nice_relations} all the above invariants can be expressed in terms of eigenvalues of matrices $W_a$, $W_s$, $W_s + i W_a$ and $W_a W_s$. Expressions~\eqref{expr} show that our approach allows to reduce the maximum power of angular coefficients entering invariants from the fifth to the fourth.

\section{\label{sec:level1}Positivity Constraints for invariants}

In order to constrain the invariants we may consider the total matrix W. Characteristic equation on its eigenvalues takes the following form:
\begin{eqnarray}\label{eq_tot_tot}
    {w}^3 - {w}^2 +\left(\frac{1}{3} \left(1-4 U_2\right)-4 U_1\right)w \nonumber \\
    -\frac{1}{27} \left(8 (R+T)-12 \left(3 U_1+U_2\right)+1\right) & = & 0
\end{eqnarray}
with
\begin{subequations}\label{EVtot_comb}
\begin{equation}
    w_1 w_2 + w_1 w_3 + w_2 w_3  = \frac{1}{3} \left(1-4 U_2\right)-4 U_1,
\end{equation}
\begin{equation}
    w_1 w_2 w_3  = \frac{1}{27} \left(8 (R+T)-12 \left(3 U_1+U_2\right)+1\right).
\end{equation}
\end{subequations}

Hadronic tensor being a product of quark currents is a semi-positive quadratic form~\cite{artru2009spin}. This means that eigenvalues are restricted to be greater or equal to zero. In addition, using the normalization condition
\begin{equation}
    \text{Tr} \hspace{3pt} W = w_1 + w_2 + w_3 = 1,
\end{equation}
we can set an upper bound on eigenvalues. Positivity and normalization together give us the following inequalities:
\begin{equation}
   0 \le w_{1, 2, 3} \le 1
\end{equation}
from which it follows that
\begin{equation}\label{cond}
\begin{gathered}    
    0 \le w_1 w_2 + w_1 w_3 + w_2 w_3 \le \frac{1}{3},\\
    0 \le w_1 w_2 w_3 \le \frac{1}{27}.
\end{gathered}
\end{equation}
Applying~\eqref{cond} to~\eqref{EVtot_comb} and using the fact that $U_1$ and $U_2$ are non-negative according to definitions~\eqref{U1} we end up with the following restrictions on introduced invariants
\begin{subequations}\label{in_U}
\begin{equation}
0 \le \frac{1}{3} - 4 U_1 - \frac{4}{3} U_2 \le 1
\end{equation}
\begin{equation}
 U_1 + \frac{1}{3} U_2\le \frac{1}{12}, \hspace{10pt} U_1 \le \frac{1}{12}, \hspace{10pt} U_2 \le \frac{1}{4}
\end{equation}
\end{subequations}

\begin{subequations}\label{in_RT}
\begin{equation}
0 \le 8 (R + T) - 12 (3 U_1 + U_2) + 1 \le 1
\end{equation}
\begin{equation}
-\frac{1}{8} \le R + T \le \frac{3}{8}
\end{equation}
\end{subequations}

\section{\label{sec:level1}Invariants for Special Rotations}
In above sections we have introduced the method which allows to find $SO(3)$ rotational invariants. One can also be interested in finding $SO(2)$ invariant quantities for rotations around a fixed axis (see e.g. \cite{Peng:2018tty}). This might be important when one wants to compare measurements done in different coordinate frames, related by special rotations. For example, three widely used in polarization experiments frames, the Helicity frame, Collins-Soper and Gottfried-Jackson frame, are related by rotation around $y$-axis~\cite{faccioli2010towards}.

Indeed, if we consider rotations around a fixed axis all the $SO(3)$ invariants we discussed before are still relevant, however additional conserving parameters appear. 

Let us consider an arbitrary vector $\bar{x}$ and a corresponding scalar $X$ of the form
\begin{equation}\label{scalar}
    X = x^{T} W x.
\end{equation}
Now, if one performs a rotation given by an arbitrary orthogonal matrix $S$:
\begin{equation}
    W^{\prime} = S^{T} W S, \hspace{10pt} x^{\prime} = S^{T} x,\hspace{10pt} {x^{\prime}}^{T} = x^{T} S,
\end{equation}
the corresponding parameter $X^{\prime}$ in the primed coordinate frame must be equal to $X$ in the unprimed one:
\begin{equation}\label{X}
    X^{\prime} = {x^{\prime}}^{T} W^{\prime} x^{\prime} = {x}^{T} S S^T W S S^T x = X,
\end{equation}
which is satisfied, since $S$ is an orthogonal matrix and $S^T = S^{-1}$. It might seem that~\eqref{X} gives a recipe for construction of an infinite number of parameters preserved by any rotation $S$, but it is not the case. Even though $X$ and $X^{\prime}$ are equal to each other they do not necessary have the same form in terms of primed and unprimed parameters of angular distribution in two different coordinate frames. For instance, consider a basis vector $e_x = (0, 1, 0)^T$, then corresponding scalar~\eqref{scalar} looks as follows:
\begin{equation}\label{I_x}
    I_x = e_x^T W e_x = \frac{1 + \lambda_\theta - 2 \lambda_\phi}{3 + \lambda_\theta}.
\end{equation}
To write~\eqref{I_x} we used the explicit form of the hadronic tensor~\eqref{h_tensor_general_lambda}. If now we consider rotation $S_y$ around unity vector $e_y = (0, 0, 1)^T$, the vector $e_x$ transforms to $e_x^\prime = (-\sin \xi, \cos \xi, 0)^T$, where $\xi$ is rotational angle. $I_x^\prime$ takes then the following form in the new frame:
\begin{equation}
    I_x^\prime = \frac{1 + \lambda_\theta^\prime \cos 2\xi + 2 \lambda_{\theta\phi}^\prime \sin 2 \xi -\lambda_\phi^\prime - \lambda_\phi^\prime \cos 2\xi}{3 + \lambda_\theta^\prime}.
\end{equation}
It is straightforward to check that $I_x = I_x^\prime$, by expressing primed parameters in terms of unprimed or vice versa. This example clearly shows that $I_x$ might have different form in different coordinate systems and thus scalars in a form~\eqref{X} are not the invariants we are looking for. However, one still can apply~\eqref{X} to construct useful invariants. To do so let us consider a special type of rotations around the $\bar{x}$ axis itself:
\begin{equation}
    W^{\prime} = S^T_x W S_x, \hspace{10pt} x^\prime = x, \hspace{10pt} {x^\prime}^T = x^T.
\end{equation}
For rotations of this type vector $x$ is preserved and
\begin{equation}
X^\prime = x^T W^\prime x = x^T W x = X,
\end{equation}
which ensures the same form of scalar $X$ in all coordinate frames related by rotation $S_x$.

It is practically important to study special rotations around coordinate axes. In our notation~\eqref{n} coordinate vectors look as follows:
\begin{equation}\label{coord_vecrots}
e_z = \left(\begin{array}{c}
        1\\
        0\\
        0\\
\end{array}\right), \hspace{10pt}
e_x = \left(\begin{array}{c}
        0\\
        1\\
        0\\
\end{array}\right), \hspace{10pt}
e_y = \left(\begin{array}{c}
        0\\
        0\\
        1\\
\end{array}\right).
\end{equation}
Contracting coordinate vectors~\eqref{coord_vecrots} with hadronic tensor~\eqref{h_tensor_general_lambda} we obtain diagonal elements of the matrix. This gives us invariant for rotations around z-axis in a form of
\begin{subequations}\label{I_}
\begin{equation}\label{z_inv}
    I_z = e_z^i W_{ij} e_z^j = \frac{1 - \lambda_\theta}{3 + \lambda_\theta}
\end{equation}
for rotations around x-axis
\begin{equation}\label{x_inv}
    I_x = e_x^i W_{ij} e_x^j = \frac{1 + \lambda_\theta - 2 \lambda_\phi}{3 + \lambda_\theta}
\end{equation}
and, finally, for rotations around y-axis
\begin{equation}\label{F}
    I_y = e_y^i W_{ij} e_y^j = \frac{1 + \lambda_\theta + 2 \lambda_\phi}{3 + \lambda_\theta}
\end{equation}
\end{subequations}
One can notice that invariance of~\eqref{z_inv} is equivalent to invariance of $\lambda_\theta$. This result and also the invariance of~\eqref{x_inv} were previously derived in~\cite{faccioli2011model}. Invariant $I_y$ is a well known parameter $\mathcal{F}$ introduced in~\cite{faccioli2010rotation} which is also related as $\mathcal{F}=(1+\lambda_0)/(3+\lambda_0)$ to the $\lambda_0$ coefficient   in the privileged frame \cite{OT05} where only polar angular distribution is present.

Note, this method also can be applied to find invariants of rotations around an arbitrary fixed direction. Let us consider rotational axis
\begin{equation}
e = \left(\begin{array}{c}
        a\\
        b\\
        c\\
\end{array}\right)
\end{equation}
then invariant would be
\begin{eqnarray}
    e^i W_{ij} e^j = a^2 I_z + b^2 I_x + c^2 I_y \nonumber \\
    -\frac{4}{3 + \lambda_\theta} \left(ab \lambda_{\theta \phi} + ac \lambda_{\perp \theta \phi} + bc\lambda_{\perp \phi}\right).
\end{eqnarray}

Another group of invariants which naturally appear when rotation around a coordinate axis is studied is a minor corresponding to this axis. For example, if we consider rotations around $z$-axis given with the matrix
\begin{equation}\label{Sez}
S_{e_z} = \left(\begin{array}{ccc}
        1 & 0 & 0\\
        0 & \cos \xi & -\sin \xi \\
        0 & \sin \xi & \cos \xi\\
\end{array}\right),
\end{equation}
where $\xi$ is a rotational angle in $xy$-plane, the invariant would be $I_{zz}$, which is equal to the determinant of the submatrix $W_{zz}$ of the matrix $W$ obtained by removing the first row and the first column.
\begin{subequations}
\begin{equation}\label{Mzz}
I_{zz} = \frac{1}{4} \left(I_z - 1 \right)^2 - \frac{4 A_{\theta}^2}{(3 + \lambda_\theta)^2} - \frac{4(\lambda_{\perp \phi}^2 + \lambda_{\phi}^2)}{(3 + \lambda_\theta)^2}
\end{equation}
The first term in~\eqref{Mzz} is invariant due to invariance of $I_z$. From the form of the considered transformation~\eqref{Sez} it follows that submatrix $W_{zz}$ transforms without mixing with the rest of $W$. The antisymmetric part of $W_{zz}$ has only one element $\frac{A_\theta}{3 + \lambda_\theta}$ and also transforms independently. That is why this combination is preserved by rotation and as a consequence the second and the third terms of~\eqref{Mzz} are both invariant under $SO(2)$ rotation around $z$-axis:
\begin{equation}\label{Izz}
    I_{zz}^{(a)} = \frac{A_\theta^2}{(3 + \lambda_\theta)^2}, \hspace{20pt} I_{zz}^{(s)} = \frac{\lambda_{\perp \phi}^2 + \lambda_{\phi}^2}{(3 + \lambda_\theta)^2},
\end{equation}
\end{subequations}
where we neglect unimportant numerical factors.

Using similar reasoning we write the determinant of matrix $W_{xx}$, which is invariant under rotation around $x$-axis:
\begin{subequations}
\begin{equation}\label{Mxx}
I_{xx} = \frac{1}{4} \left(I_x - 1 \right)^2 - 4 I_{xx}^{(a)} - I_{xx}^{(s)},
\end{equation}
where
\begin{equation}\label{Ixx}
    I_{xx}^{(a)} = \frac{A_\phi^2}{(3 + \lambda_\theta)^2}, \hspace{20pt} I_{xx}^{(s)} = \frac{4 \lambda_{\perp\theta\phi}^2 + (\lambda_\theta + \lambda_\phi)^2}{(3 + \lambda_\theta)^2}
\end{equation}
\end{subequations}
are also $SO(2)$ invariants.

For rotations around $y$-axis we obtain invariant minor
\begin{subequations}
\begin{equation}\label{Myy}
I_{yy} = \frac{1}{4}\left( I_y - 1\right)^2 - 4 I_{yy}^{(a)} - I_{yy}^{(s)},
\end{equation}
with invariants
\begin{equation}\label{Iyy}
I_{yy}^{(a)} = \frac{A_{\perp \phi}^2}{(3 + \lambda_\theta)^2}, \hspace{20pt} I_{yy}^{(s)} = \frac{4\lambda_{\theta \phi}^2 + (\lambda_\theta - \lambda_\phi)^2}{(3 + \lambda_\theta)^2}.
\end{equation}
\end{subequations}
Invariant $I_{yy}^{(s)}$ were previously derived in~\cite{palestini2011angular} and then rederived in~\cite{ma2017rotation}.

Note, that because of the positivity and normalization conditions invariants~\eqref{I_},~\eqref{Mzz},~\eqref{Mxx} and~\eqref{Myy} can be restricted:
\begin{subequations}\label{in_special}
\begin{equation}
	0 \le I_z, I_x, I_y \le 1,
\end{equation}
\begin{equation}
	0 \le I_{zz}, I_{xx}, I_{yy} \le \frac{1}{4}.
\end{equation}
\end{subequations}
The last inequality is relevant because $I_{zz}$, $I_{xx}$, $I_{yy}$ are minors and thus are equal to the product of two corresponding eigenvalues, which are bound by normalization condition and positivity as we discussed above.

\section{\label{sec:level1}Calculation of Rotational-Invariant Parameters for $Z$ - decays}
In this section we apply the derived invariants in the form of~\eqref{notation} to the analysis of experimental results presented by ATLAS collaboration~\cite{ATLAS}. The paper presents a measurement of the full set of eight coefficients using charged lepton pairs (electrons or muons). The measurement is performed in the $Z$-boson mass peak. The data is presented as a function of $Z$-boson transverse momentum $p_T^Z$ for integrated rapidity of $Z$-boson $y^Z$ and for three bins of $y^Z$: $0 < |y^Z| < 1$, $1 < |y^Z| < 2$, $2 < |y^Z| < 3.5$. Measurement is performed in the Collins-Soper reference frame.

ATLAS uses parametrization~\eqref{general_A}. Applying substitution~\eqref{A-l_relation} to invariants~\eqref{notation} and also omitting unimportant common factors which appear as a result of parametrization change
\begin{eqnarray}\label{subst}
U_1 \rightarrow 64 U_1, \qquad U_2 \rightarrow 64 U_2, \qquad T \rightarrow 256 T, \nonumber \\* 
R \rightarrow \frac{512}{9}R, \qquad M \rightarrow 4096 M
\end{eqnarray}
we obtain the following form of invariants:
\begin{subequations}\label{Inv_A}
\begin{equation}\label{U_1_A}
	U_1 = A_3^2+A_4^2+A_7^2
\end{equation}
\begin{equation}\label{U_2_A}
	U_2 = 9 A_0^2-12 A_0+12 A_1^2+3 A_2^2+12 A_5^2+12 A_6^2+4
\end{equation}
\begin{widetext}
\begin{eqnarray}\label{T_A}
	T & = \left. 27 A_0^3-54 A_0^2+9 \left(6 A_1^2-3 A_2^2-12 A_5^2+6 A_6^2+4\right) A_0+18 A_2^2+72 A_5^2+54 A_2 A_6^2-36 A_6^2 \right.\nonumber \\
	& \left.-18 A_1^2 \left(3 A_2+2\right)-216 A_1 A_5 A_6-8\right.
\end{eqnarray}
\begin{equation}\label{R_A}
	R = \left. \left(3 A_0+3 A_2-2\right) A_3^2+12 \left(A_1 A_4+A_5 A_7\right) A_3+\left(4-6 A_0\right) A_4^2+\left(3 A_0-3 A_2-2\right) A_7^2+12 A_4 A_6 A_7 \right.
\end{equation}
\begin{eqnarray}\label{M_A}
M & = \left. -2 \left(9 A_0^2-3 \left(3 A_2+4\right) A_0+18 A_6^2+6 A_2+4\right) A_3^2-12 \left(A_1 \left(\left(3 A_0-3 A_2-2\right) A_4-6 A_6 A_7\right) \right. \right. \nonumber\\
& - \left. \left. 2 A_5 \left(3 A_4 A_6+\left(3 A_0-2\right) A_7\right)\right) A_3-2 \left(9 A_0^2+3 \left(3 A_2-4\right) A_0+18 A_1^2-6 A_2+4\right) A_7^2 \right.\nonumber \\
& + \left. A_4^2 \left(9 A_0^2-12 A_0-9 A_2^2-36 A_5^2+4\right)+12 A_4 \left(6 A_1 A_5+\left(-3 A_0-3 A_2+2\right) A_6\right) A_7\right.
\end{eqnarray}
\end{widetext}
\end{subequations}
One can also suggest to get rid of constant terms which appear in $U_2$ and $T$, but we prefer to preserve them, since this form of invariants ensures that they are all going to zero in case of isotropic distribution. This is not an essential requirement, but just a nice way to normalize introduced parameters.

Tables~\ref{tab: ATLAS_integrated} -- \ref{tab: ATLAS_y23} present the values of invariants $U_1$, $U_2$, $T$, $R$, $M$ calculated from the angular coefficients measured by ATLAS. One can note that despite the fact that invariants are polynomials of angular coefficients and thus error should add up, the results are still quite precise. This is because the data shows significant dominance of some coefficients over other and large coefficients which contribute the most to the invariants are measured with great accuracy.

\begin{table*}
\caption{\label{tab: ATLAS_integrated}The values of invariants calculated for angular coefficients measured by ATLAS collaboration~\cite{ATLAS} in the $Z/\gamma^* \rightarrow e^+e^-$ and $Z/\gamma^* \rightarrow \mu^+\mu^-$ $y^Z$-integrated channel at low ($5-8$ GeV), mid ($22-25.5$ GeV) and high ($132-173$ GeV) $p_T^Z$. The uncertainties include both statistical and systematic errors.}
\begin{ruledtabular}
\begin{tabular}{cccccc}
 $p_T$ [GeV/c] & $U_ 1 $ & $U_ 2 $ & $T$ & $R$ & $M$\\ \hline
 $5.0 - 8.0$ & $0 .0067\pm 0.0004 $ & $3 .82\pm 0.09 $ & $ - 
      7.48\pm 0.25 $ & $0 .026\pm 0.001 $ & $ 0.0064 \pm 0.0004 $ \\

$22.0 - 25.5$ & $0 .0043\pm 0.0004 $ & $ 2.37\pm 0.07 $ & $ - 
      3.45\pm 0.16 $ & $0 .013\pm 0.001 $ & $ 0.0024 \pm 0.0002 $ \\

$132 - 173$ & $0 .0037\pm 0.0009 $ & $ 1.88\pm 0.13 $ & $ - 
      2.50\pm 0.25 $ & $0 .008\pm 0.002 $ & $ 0.0010 \pm 0.0005 $ \\
\end{tabular}
\end{ruledtabular}
\end{table*}

\begin{table*}
\caption{\label{tab: ATLAS_y01}The values of invariants calculated for angular coefficients measured by ATLAS collaboration~\cite{ATLAS} in the $Z/\gamma^* \rightarrow e^+e^-$ and $Z/\gamma^* \rightarrow \mu^+\mu^-$ channels for $0 < |y^Z| < 1$ at low ($5-8$ GeV), mid ($22-25.5$ GeV) and high ($132-173$ GeV) $p_T^Z$. The uncertainties include both statistical and systematic errors.}
\begin{ruledtabular}
\begin{tabular}{cccccc}
$p_T$ [GeV/c] & $U_ 1 $ & $U_ 2 $ & $T$ & $R$ & $M$ \\ \hline
$5.0-8.0$ & $0.0010\pm 0.0001$ & $3.74\pm 0.06$ & $-7.23\pm 0.17$ & $0.0039\pm 0.0006$ & $ 0.0009 \pm 0.0001$ \\

$22.0-25.5$ & $0.0003\pm 0.0001$ & $2.38\pm 0.05$ & $-3.46\pm 0.12$ & $0.0008\pm 0.0003$ & $0.00013\pm 0.00006$ \\

$132-173$ & $0.0006\pm 0.0005$ & $1.70\pm 0.15$ & $-2.13\pm 0.27$ &$ 0.0008\pm 0.0011$ & $-0.00002\pm 0.00025 $\\
\end{tabular}
\end{ruledtabular}
\end{table*}

\begin{table*}
\caption{\label{tab: ATLAS_y12}The values of invariannts calculated for angular coefficients measured by ATLAS collaboration~\cite{ATLAS} in the $Z/\gamma^* \rightarrow e^+e^-$ and $Z/\gamma^* \rightarrow \mu^+\mu^-$ channels for $1 < |y^Z| < 2$ at low ($5-8$ GeV), mid ($22-25.5$ GeV) and high ($132-173$ GeV) $p_T^Z$. The uncertainties include both statistical and systematic errors.}
\begin{ruledtabular}
\begin{tabular}{cccccc}
$p_T$ [GeV/c] & $U_ 1 $ & $U_ 2 $ & $T$ & $R$ & $M$ \\ \hline
$5.0-8.0$ & $0.0043\pm 0.0003$ & $3.80\pm 0.09$ &$ -7.41\pm 0.28$ & $0.016\pm 0.001$ & $ 0.004 \pm 0.0003$ \\

$22.0-25.5$ & $0.0033\pm 0.0004 $& $2.48\pm 0.07$ & $-3.71\pm 0.17$ & $ 0.010\pm 0.001$ & $0.0019\pm 0.0003$ \\

$132-173$ & $0.0067\pm 0.0021 $& $1.58\pm 0.21 $& $-0.90\pm 0.19 $&$ 0.015\pm 0.005$ & $0.0007\pm 0.0007 $\\
\end{tabular}
\end{ruledtabular}
\end{table*}

\begin{table*}
\caption{\label{tab: ATLAS_y23}The values of invariants calculated for angular coefficients measured by ATLAS collaboration~\cite{ATLAS} in the $Z/\gamma^* \rightarrow e^+e^-$ channel for $2 < |y^Z| < 3.5$ at low ($5-8$ GeV), mid ($22-25.5$ GeV)$p_T^Z$. The uncertainties include both statistical and systematic errors.}
\begin{ruledtabular}
\begin{tabular}{cccccc}
$p_T$ [GeV/c] & $U_ 1 $ & $U_ 2 $ & $T$ & $R$ & $M$ \\ \hline
$5.0-8.0$ & $0.020\pm 0.003$ &$ 3.2\pm 0.4 $& $-5.6\pm 1.1$ & $0.07\pm 0.01$ &$ 0.016 \pm 0.003 $\\

$22.0-25.5$ &$ 0.014\pm 0.004$ & $2.7\pm 0.4$ & $-3.8\pm 1.0 $& $0.04\pm 0.01$ & $ 0.006 \pm 0.003 $\\
\end{tabular}
\end{ruledtabular}
\end{table*}

We rewrite inequlites~\eqref{in_U} and~\eqref{in_RT} obtained earlier taking into account the change~\eqref{subst}.
\begin{subequations}
\begin{equation}
U_1 + \frac{1}{3} U_2 \le \frac{16}{3}, \hspace{10pt} U_1 \le \frac{16}{3}, \hspace{10pt} U_2 \le 16
\end{equation}
\begin{equation}
-64 \le 9R + 2T -12 (3 U_1 + U_2) \le 0, \hspace{10pt} -64 \le 9R + 2T \le 192
\end{equation}
\end{subequations}
As one can see they are satisfied for all the invariants represented in Tables~\ref{tab: ATLAS_integrated} --~\ref{tab: ATLAS_y23}.

\section{\label{sec:level1}Calculation of Rotational-Invariant Parameters for $J/\psi$ decays}
PHENIX collaboration~\cite{adare2017angular} reports the measurement of the angular distribution for $J/\psi \rightarrow \mu^-\mu^-$ decays in $pp$-collisions. Data are available for transverse momenta $2 < p_T < 10$ GeV and rapidity $1.2 < y < 2.2$. This particular paper is especially interesting because measurements were performed in four different reference frames: the Helicity frame (HX), Collins-Soper (CS), Gottfried-Jackson Backward (GJB) and Gottfried-Jackson Forward (GJF). PHENIX uses the angular distribution in the form of~\eqref{general_lambda}, but the collaboration reports the measurement of only three coefficients $\lambda_\theta$, $\lambda_\phi$ and $\lambda_{\theta \phi}$. One can do the following: assume the remaining coefficients to be equal to zero, calculate invariants for different coordinate frames and then if there is inconsistency between the results which can not be explained by statistical and systematic errors make predictions about values of the coefficients which were not measured in the experiment.

Let us work with the set of invariants~\eqref{notation}. We are left with only two non zero parameters $U_2$ and $T$ if zero values of $\lambda_{\perp \phi}$, $\lambda_{\perp\theta\phi}$, $A_\phi$, $A_\theta$ and $A_{\perp \phi}$ are assumed. $U_2$ and $T$ can potentially give us information about two not measured coefficients $\lambda_{\perp\phi}$ and $\lambda_{\perp\theta\phi}$. Table~\ref{tab: J/psi} shows the values of invariants for angular coefficients in $J/\psi \rightarrow \mu^-\mu^-$ decays measured by PHENIX collaboration. As one can see data is consistent with the assumption of zero values of coefficients $\lambda_{\perp \phi}$ and $\lambda_{\perp \theta \phi}$.

\begin{table*}
\caption{\label{tab: J/psi}The values of invariants $U_2$ and $T$ calculated for angular coefficients measured by PHENIX collaboration~\cite{adare2017angular} in $J/\psi \rightarrow \mu^-\mu^-$ decays for $1.2 < y < 2.2$ in four reference frames at different values of $p_T$: ($2-3$ GeV), ($3-4$ GeV) and ($4-10$ GeV). Only statistical errors are taken into account.}
\begin{ruledtabular}
\begin{tabular}{ccccccc}
 & \multicolumn{3}{c}{$U_2$} & \multicolumn{3}{c}{$T$} \\
$p_T$ [GeV/c] & $2-3$ & $3-4 $& $4-10$ & $2-3$ &$ 3-4 $&$ 4-10$ \\
\hline
HX & $3.0\pm 2.5$ & $2.8\pm 1.7$ & $1.2\pm 0.8 $ & $4.8\pm 6.5$ & $1.7\pm 3.7$ & $1.3\pm 1.4$ \\
CS & $> (6.2\pm 0.4)$ & $1.0\pm 0.8$ & $0.5\pm 0.8$ &$ > (15.2\pm 1.5)$ & $-0.1\pm 1.6$ & $-0.3\pm 0.8 $\\
GJB & $5.0\pm 3.6 $& $3.8\pm 3.6 $& $1.0\pm 0.6$ & $10.5\pm 12.1 $& $0.7\pm 6.9$ &$ 0.9\pm 0.8$ \\
GJF & $8.7\pm 4.4$ & $3.0\pm 1.7$ & $4.3\pm 2.2$ & $24.3 \pm 20.1$ & $3.3\pm 3.5$ & $7.4\pm 7.1$ \\
\end{tabular}
\end{ruledtabular}
\end{table*}

PHENIX collaboration also presents the calculation of the $y$-rotation invariant angular parameter
\begin{equation}\label{lambda_tilda}
	\tilde{\lambda} = \frac{\lambda_\theta + 3 \lambda_\phi}{1 - \lambda_\phi}.
\end{equation}
This parameter despite being sensitive to the maximum angular asymmetry~\cite{faccioli2010towards} also has another interpretation. Let us assume that there exist a frame where angular distribution takes the following form with respect to some axis:
\begin{equation}\label{geometric_model}
        \frac{1}{\sigma}\frac{\text{d} \sigma}{\text{d}\Omega} = \frac{3}{4 \pi} \frac{1}{3 + \lambda_\theta} \left(1 + \lambda_0 \cos^2 \theta\right).
\end{equation}
One can perform a rotation around $y$-axis with rotational angle $\xi$. This lead to distribution in the form used by PHENIX:
\begin{eqnarray}\label{dist_PHENIX}
\frac{1}{\sigma}\frac{\text{d} \sigma}{\text{d}\Omega} & = & \frac{3}{4 \pi} \frac{1}{3 + \lambda_0} \left(1 + \lambda_\theta \cos^2 \theta \right.\nonumber \\
& + & \left. \lambda_{\theta \phi} \sin 2 \theta \cos \phi + \lambda_\phi \sin^2 \theta         \cos 2\phi\right).
\end{eqnarray}
As it was shown in~\cite{OT05,teryaev2011positivity} then parameters $\lambda_0$ and $\sin^2 \xi$ can be written through angular coefficients of distribution~\eqref{dist_PHENIX}
\begin{subequations}\label{lambda0_xi}
\begin{equation}\label{lambda_0}
        \lambda_0 = \frac{\lambda_\theta + 3 \lambda_\phi}{1 - \lambda_\phi}
\end{equation}
\begin{equation}\label{xi}
        \sin^2 \xi = \frac{2 \lambda_\phi}{\lambda_\theta + 3 \lambda_\phi}
\end{equation}	
\end{subequations}
Comparing~\eqref{lambda_tilda} and~\eqref{lambda_0} we see that invariant parameter $\tilde{\lambda}$ receives new interpretation as an angular coefficient in front of $\cos^2 \theta$ in the reference frame where distribution has azimuthally symmetric form~\eqref{geometric_model}, while~\eqref{xi} gives sine squared of the angle which relates the frame with angular distribution~\eqref{dist_PHENIX} and the frame with distribution~\eqref{geometric_model}. 

\begin{figure}[h]
        \includegraphics[width=0.65\linewidth]{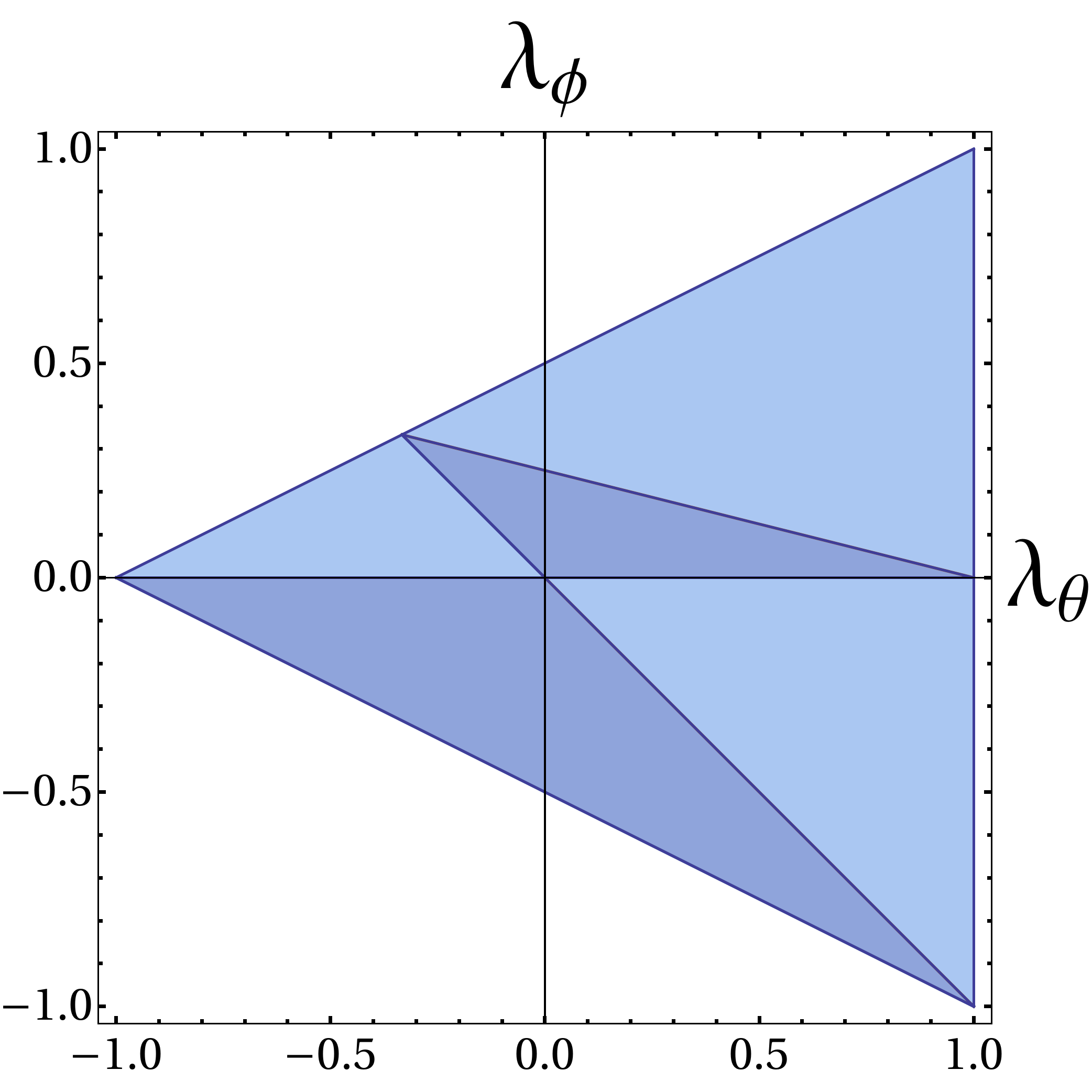}
        \caption{\label{fig: domains}Allowed domains for parameters $\lambda_\theta$ and $\lambda_\phi$. Greater triangle with vertexes $(-1, 0)$, $(1,1)$, $(1,-1)$ corresponds to the region allowed by positivity conditions~\eqref{positivity_cond}. Smaller shadowed triangles correspond to points in parameter space for which there exist a frame where distribution takes azimuthally symmetric form.}
\end{figure}

However, in general, such a frame not necessary exist. First, positivity conditions restrict $\lambda_0$ to take its values between $-1$ and $1$. Secondly, $\sin^2 \xi$ can vary only between $0$ and $1$. Thus from~\eqref{lambda0_xi} we obtain restrictions on angular parameters $\lambda_\theta$ and $\lambda_\phi$:
\begin{subequations}\label{ineq_l_xi}
\begin{equation}
-1 \le \frac{\lambda_\theta + 3 \lambda_\phi}{1 - \lambda_\phi} \le 1
\end{equation}
\begin{equation}\label{positivity_cond}
0 \le \frac{2 \lambda_\phi}{\lambda_\theta + 3 \lambda_\phi} \le 1.
\end{equation}
\end{subequations}
If these inequalities are satisfied for the angular distribution in a form of~\eqref{dist_PHENIX}, there exist a frame  where distribution is azimuthally symmetric with parameter $\lambda_0$ given by~\eqref{lambda_0}. This frame is related to the frame under consideration by rotation around $y$ axis by the angle given in~\eqref{xi}. Note, that positivity conditions written for parameters $\lambda_\theta$ and $\lambda_\phi$ allow to violate~\eqref{ineq_l_xi}. Fig.~\ref{fig: domains} shows two regions in parameter plane $(\lambda_\theta, \lambda_\phi)$. The bigger shadowed triangle with vertexes $(-1, 0)$, $(1, 1)$, $(1, -1)$ corresponds to allowed values of parameters according to positivity conditions:
\begin{equation}
|\lambda_\theta| \le 1, \hspace{10pt} |2\lambda_\phi| \le 1 + \lambda_\theta.
\end{equation}
Two smaller dark triangles are regions truncated by inequalities~\eqref{ineq_l_xi}. For those values of parameters $\lambda_\theta$ and $\lambda_\phi$, which belong to truncated triangles one can find a coordinate system where angular distribution is azimuthally symmetric with parameters~\eqref{lambda0_xi}.

\begin{figure}[h]
		\centering 
		\subfigure[]{
			\includegraphics[width=0.45\linewidth]{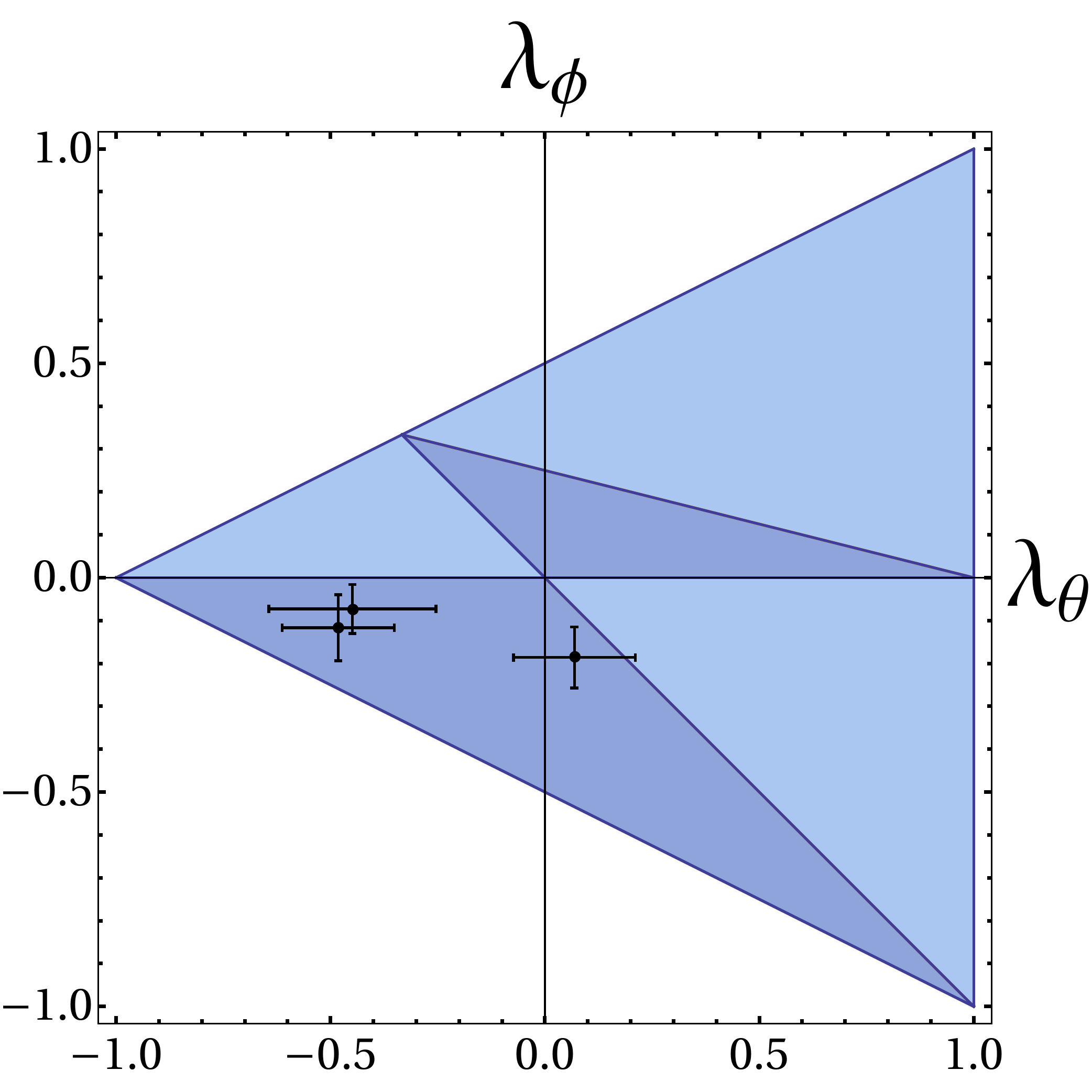}
			\label{im:HX}}
		\subfigure[]{
			\includegraphics[width=0.45\linewidth]{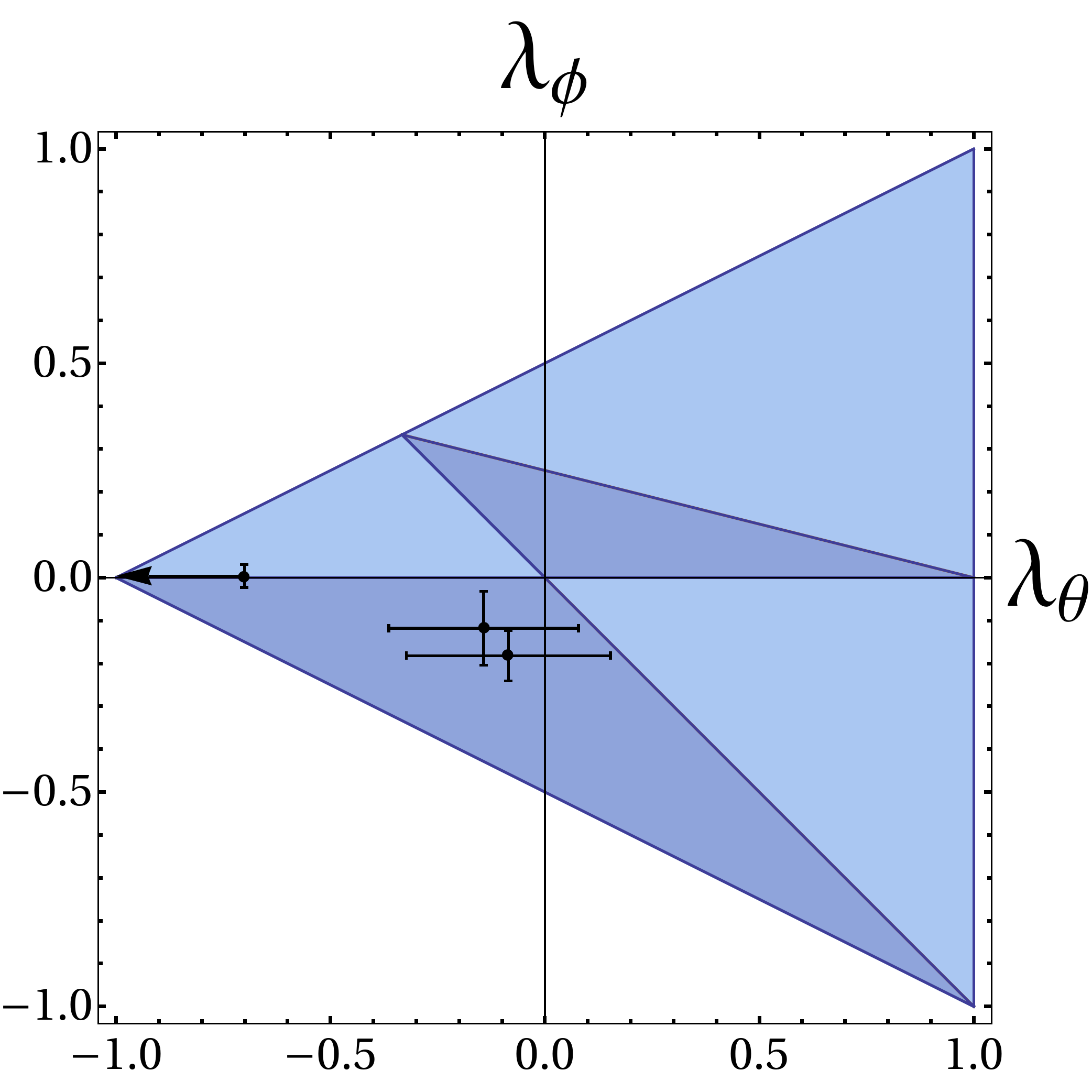}
			\label{im:CS}}
			\\
					\subfigure[]{
			\includegraphics[width=0.45\linewidth]{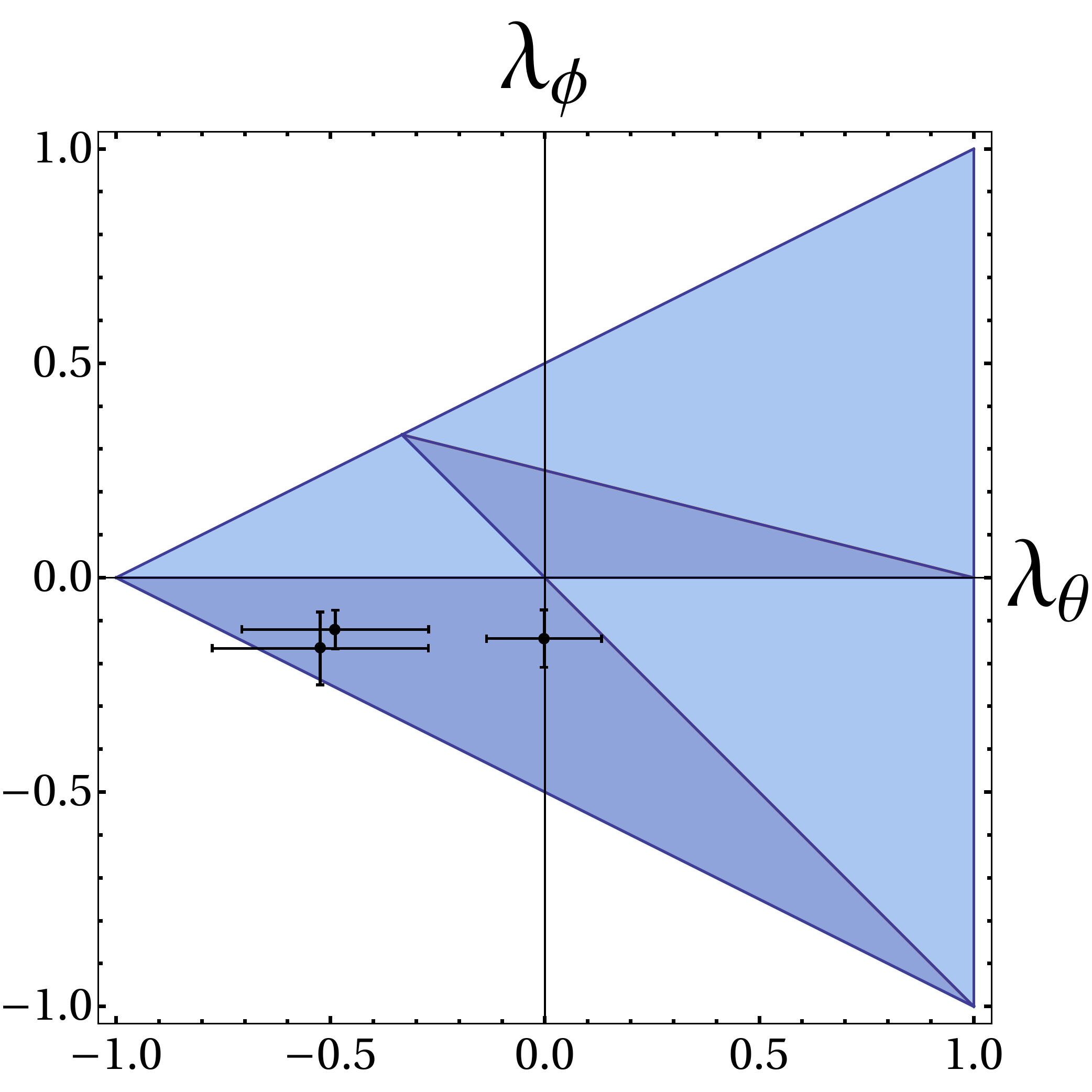}
			\label{im:GJB}}
		\subfigure[]{
			\includegraphics[width=0.45\linewidth]{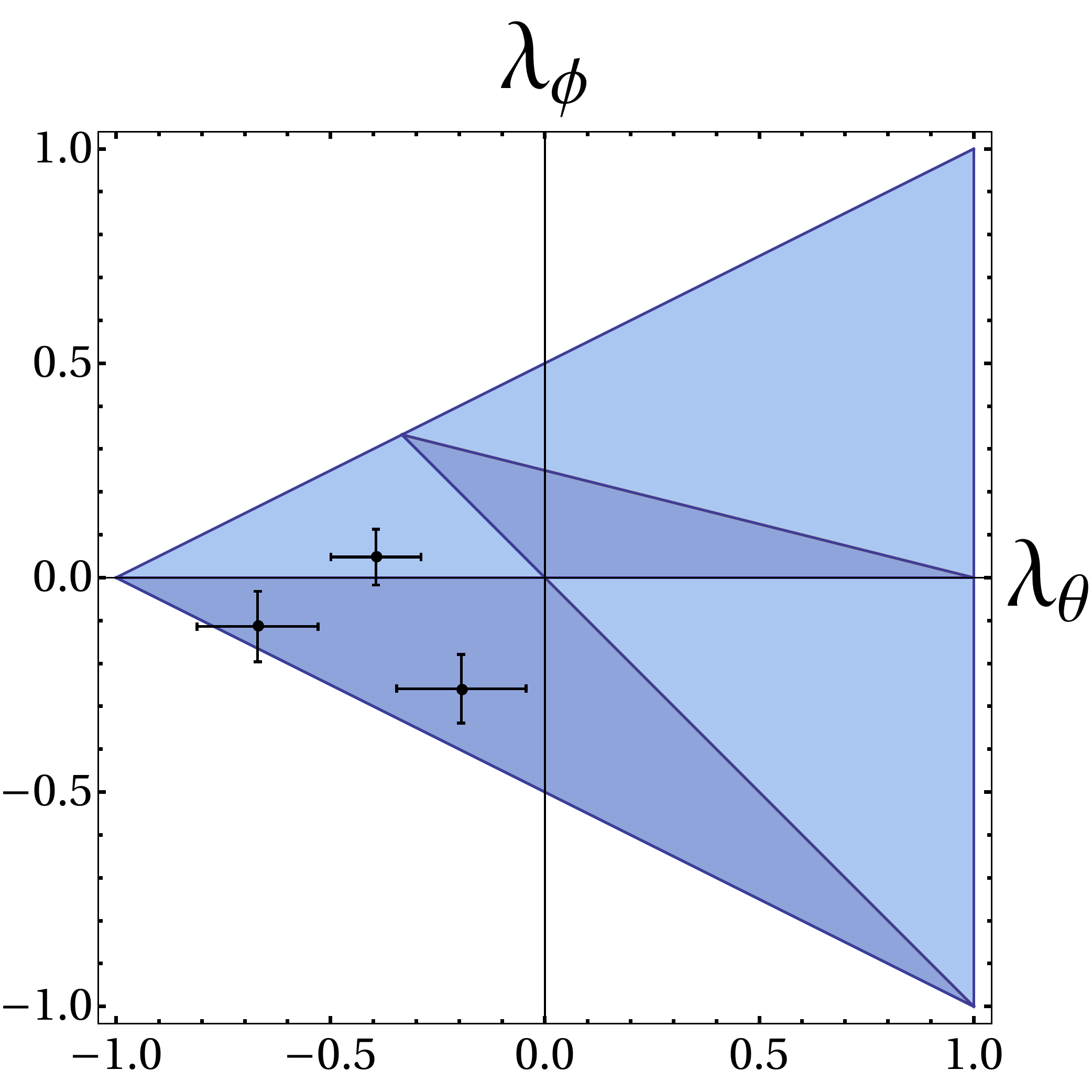}
			\label{im:GJF}}
		\caption{\label{im:PHENIX}Angular coefficients $\lambda_\theta$ and $\lambda_\phi$ measured by PHENIX~\cite{adare2017angular}: \subref{im:HX} -- HX frame, \subref{im:CS} -- CS frame, \subref{im:GJB} -- GJB frame, \subref{im:GJF} -- GJF frame. Different points correspond to different values of transverse momentum. Only statistical errors are shown.}
	\end{figure}

Figure~\ref{im:PHENIX} shows angular coefficients measured in $J/\psi$ decays plotted on $(\lambda_\theta, \lambda_\phi)$ plane. As one can see all the data points belong to regions where parameters~\eqref{lambda0_xi} exist.

\section{\label{sec:level1}Conclusions}
Experimental studies of vector decays into fermion pairs are usually conducted by measuring the coefficients of the angular distribution of final state particles. In this work, we have shown that all the information about distribution can be expressed in a form of a single matrix~\eqref{h_tensor_general_lambda}. However, according to~\eqref{rotation_S} its elements depend on the choice of a coordinate system. That is why rotational-invariant combinations of angular coefficients are expected to be better observables. A bunch of such invariants for special cases was introduced in literature~\cite{faccioli2010rotation, faccioli2010rotationPV, palestini2011angular, faccioli2011model}, also a general method for their derivation was recently proposed~\cite{ma2017rotation}. 

In our work we developed formalism which allowed us to find a set of five $SO(3)$ rotational invariants~\eqref{notation} and relate them to $SO(3)$ invarinat parameters obtained earlier~\eqref{expr} in the work~\cite{ma2017rotation}. The significant feature of the set of invariants that we propose is their more compact form and, in particular, the reduced maximum power of the angular coefficients entering invariants.  We have also shown how the developed formalism can be used for derivation of invariants for special rotations around fixed axes and reproduced previous results~\eqref{I_},~\eqref{Iyy}. Additionally, we have found two pairs of special invariants~\eqref{Izz} and~\eqref{Ixx} which for our best knowledge were not presented in literature before. Moreover, the hadronic tensor formalism allowed as to constrain some of $SO(3)$ frame independent parameters~\eqref{in_U},~\eqref{in_RT} and also invariants for special rotations~\eqref{in_special}.

In two later sections we have calculated invariants for experimental data. Tables~\ref{tab: ATLAS_integrated}--\ref{tab: ATLAS_y23} show results for $Z$ decays, Table~\ref{tab: J/psi} and Figure~\ref{im:PHENIX} summarize results for $J/\psi$ decays.

By the time when the current work was essentially done, we became aware of the very similar recent interesting work on the topic~\cite{martens2018quantum}. The two works are using basically the same density matrix (hadronic tensor) approach to the dilepton angular distribution studies, but they also have some differences. First, we explicitly present the expression for the density matrix in terms of angular coefficients. Second, we use different sets of invariants. Authors of the work~\cite{martens2018quantum} suggest to use eigenvalues of matrices $W_s$, $W_a$ and scalar products of the vector part of the density matrix and eigenvectors of the symmetric part as invariant parameters. In our analysis we considered eigenvalues of $W_s$ and $W_a$ as well as eigenvalues of matrices $W_s + i W_a$ and $W_a W_s$. We have explicitly written the expressions for eigenvalues (Appendix) and suggested to use more convenient invariants listed in~\eqref{notation}. The more detailed comparison of approaches is still of interest. 

\begin{acknowledgments}
We are most indebted to Jen-Chieh Peng, Wen-Chen Chang and Randall Evan McClellan for the very useful discussions and valuable comments. O.T. is also thankful to Alexander Artamonov, Daniel Boer and Jian-Wei Qiu for interesting and inspiring discussions and comments and to John Ralston for the very helpful correspondence and deep remarks. 

The work was partially supported by RFBR Grant 17-02-01108.
\end{acknowledgments}

\begin{widetext}
\appendix*
\section{Eigenvalues}
\begin{subequations}\label{roots}
\begin{equation}\label{EVa}
w_1^{(a)} = 0, \hspace{10pt} w_{2, 3}^{(a)} = \pm 2 \sqrt{-U_1}
\end{equation}
\begin{equation}\label{EVs1}
w_1^{(s)} = \frac{2^{2/3} \left(\sqrt{T^2-4 U_2^3}+T\right){}^{2/3}+2^{4/3} U_2}{3 \sqrt[3]{\sqrt{T^2-4 U_2^3}+T}}
\end{equation}
\begin{equation}\label{EVs23}
w_{2,3}^{(s)} = -\frac{\sqrt[3]{2} \left(\sqrt{T^2-4 U_2^3}+T\right){}^{2/3}+2 U_2}{3 \cdot 2^{2/3} \sqrt[3]{\sqrt{T^2-4 U_2^3}+T}} \pm \frac{i \left(2^{2/3} \left(\sqrt{T^2-4 U_2^3}+T\right){}^{2/3}-2^{4/3} U_2\right)}{2 \sqrt{3} \sqrt[3]{\sqrt{T^2-4 U_2^3}+T}}
\end{equation}
\begin{equation}\label{EVtot1}
w_1 = \frac{\sqrt[3]{2} \left(\sqrt[3]{2} \left(\sqrt{(R+T)^2-4 \left(3 U_1+U_2\right){}^3}+R+T\right){}^{2/3}+6 U_1+2 U_2\right)}{3 \sqrt[3]{\sqrt{(R+T)^2-4 \left(3 U_1+U_2\right){}^3}+R+T}}
\end{equation}
\begin{eqnarray}\label{EVtot23}
w_{2,3} = -\frac{\sqrt[3]{2} \left(\sqrt{(R+T)^2-4 \left(3 U_1+U_2\right){}^3}+R+T\right){}^{2/3}+6 U_1+2 U_2}{3 \cdot 2^{2/3} \sqrt[3]{\sqrt{(R+T)^2-4 \left(3 U_1+U_2\right){}^3}+R+T}} \nonumber \\
-\frac{i \left(\sqrt[3]{2} \left(\sqrt{(R+T)^2-4 \left(3 U_1+U_2\right){}^3}+R+T\right){}^{2/3}-6 U_1-2 U_2\right)}{2^{2/3} \sqrt{3} \sqrt[3]{\sqrt{(R+T)^2-4 \left(3 U_1+U_2\right){}^3}+R+T}}
\end{eqnarray}
\begin{equation}\label{EVas}
w_1^{(as)} = 0, \hspace{10pt} w_{2,3}^{(as)} = \pm \frac{4}{3} \sqrt{-M}
\end{equation}
\end{subequations}
\end{widetext}

\bibliography{RIO_bib}

\end{document}